\documentstyle[aps,prl,floats,epsfig]{revtex}

\begin{document}
\draft

\twocolumn[\hsize\textwidth\columnwidth\hsize\csname
@twocolumnfalse\endcsname

\title{Current-driven magnetization decrease in single crystalline ferromagnetic manganese oxide }
\author{A. Wahl, Ch. Simon, S. Mercone, D. Saurel, C. Martin }
\address{Laboratoire CRISMAT, Unit\'{e} Mixte de Recherches 6508, Institut des\\
Sciences de la Mati\`{e}re et du Rayonnement - Universit\'{e} de Caen, 6\\
Boulevard du\\
Mar\'{e}chal Juin, 14050 Caen Cedex, France.}
\date{\today}
\maketitle

\begin{abstract}
The electrical and magnetic response to a bias current has been investigated 
in a singlecrystalline ferromagnetic manganese oxide $\Pr_{0.8}$Ca$_{0.2}$MnO$_3$ . 
A significant decrease of the magnetization 
is observed at the same threshold current where a non-linearity of V-I characteristics 
appears. Such a behavior cannot be understood in the framework of the filamentary picture 
usually invoked for the non linearity of the other manganese oxides. Instead, an analogy with spintronic features 
might be useful and experimental signatures seem to be in agreement with excitations of 
spin waves by an electric current. This provides an example of a bulk system in which the spin polarized current induces 
a macroscopic change in the magnetization.
\end{abstract}

]

The physics of the phase transition from a charge-ordered antiferromagnet
(CO-AF) to a charge delocalized ferromagnet (CD-FM) in bulk manganese oxides
has recently been under very active investigation.\cite{REVIEW}\ This CO
destabilization is of great interest because this feature can be achieved
under a wide variety of external perturbations.\cite{LEE95,KIR97,OGA98} For
instance, numerous experimental results have shown that application of a
moderate electric field leads to a metal-insulator (MI) transition
associated with a strong non-linearity of V-I characteristics.\cite
{ASA97,STA00,GUH00ab,RAO00}\ This MI transition is not believed to affect
the entire bulk of the sample: conducting filaments along the current path,
stabilized by FM correlations (due to DE mechanism), may occur instead, and
a gradual melting of the CO-AF phase would follow. This scenario increases
considerably the technological potential of the AF-CO manganese oxides in
the form of thin films, since nanoscale production of heterostructure
geometry is a key requirement in all technologies. Recently, we have
observed the same kind of feature in a non charge-ordered ferromagnetic
insulator bulk crystal, i.e. a strong drop in resistance under application
of a moderate electric field concomitant with non-linearity in V-I
characteristics.\cite{MER02} However, the physics at work here are quite
different than those in the former case, because the CO ground state can not
be invoked as a key ingredient for the understanding of this feature.

Among various manganites, the $\Pr_{1-x}$Ca$_x$MnO$_3$ system (PCMO) is
unique since it shows insulating behavior over the whole composition range
due to the narrow bandwidth of the $e_g$ electrons.\cite{MAR99,JIR85} Within
the doping range $0.3\leq x\leq 0.5$, the ground state of the PCMO system is
a CO-AF insulator; the real space ordering of 1:1 $Mn^{3+}$/ $Mn^{4+}$ ions
occurs at 270K, and the local spin moments order at 170K.\ At lower doping
levels ($x\leq 0.25)$, this system exhibits a non trivial FM insulator
ground state. In this letter, we address the specific question of the
current-induced excitation of the magnetic state of an insulating
ferromagnet and its connection with the non-linearity of the electric
response. To do so, we have examined simultaneously the electrical and
magnetic response to a bias current of a non charge-ordered bulk crystal ($%
\Pr_{0.8}$Ca$_{0.2}$MnO$_3$). In order to put our results in perspective,
data obtained for a CO compound ($\Pr_{0.63}$Ca$_{0.37}$MnO$_3$) are also
presented.

For this study, single crystals of $\Pr_{0.8}$Ca$_{0.2}$MnO$_3$ and $%
\Pr_{0.63}$Ca$_{0.37}$MnO$_3$ were grown using the floating zone method in
an image furnace. Experimental details are presented elsewhere, including
electron diffraction and spectroscopic analysis.\cite{HAR02,WAHL02} The V-I
characteristics and magnetization versus current curves were measured
simultaneously by means of a SQUID\ magnetometer and an external current -
voltage source (Keithley 236).

Figure 1 shows the V-I characteristics for $\Pr_{0.8}$Ca$_{0.2}$MnO$_3$
(Fig. 1b) and $\Pr_{0.63}$Ca$_{0.37}$MnO$_3$ (Fig. 1a) crystals. Non-linear
effects are observed for both

compounds with the occurrence of negative differential resistance when the
bias current attains a current threshold $(J_{th})$. In both cases, the V-I
characteristics are not hysteretic. These two materials, whose electrical
ground state is insulating, are thus driven to become conducting when a
sufficient bias current is attained. Such a result is now well-established
for the CO-AF ground state and its occurrence for low-doped FM insulator
manganites suggests that this feature is not linked to this specific ground
state. One may speculate that the current induced delocalization of carriers
may follow a different process for the FM and CO-AF compounds.\cite{MER02}
We have carefully determined that the Joule heating does not account for
this current-induced effect.\ The temperature rise of the sample with
respect to the sample holder $\left( \Delta T\right) $ was measured by
attaching a thermometer to the top of the sample itself. In this low
temperature range, the power dissipation level where the voltage drop sets
in leads to $\Delta T\prec 3K$. For higher temperatures, $\Delta T$ becomes
negligible. Moreover, the same non linear effects are produced when keeping
the current in the sample constant and reducing the cross sectional area of
the sample by a factor of ten, thus increasing the current density, which is
the controlling factor of such behavior.

The magnetic signature of the current-driven MI transition is of great
interest to understand the nature of the interaction between the current and
the local magnetic moments. Concerning the CO-AF system and within the
framework of the CO phase destabilization, there should be a magnetic
signature of the transition in terms of an enhanced magnetic moment.\cite
{STA00} This is confirmed in Fig. 1a where a significant rise in
magnetization is observed at the same current threshold as the voltage
drops. We obtain a 17\% rise in magnetization which seems to fit the
filamentary picture. However, although the electrical response is identical
for the FM and CO-AF ground states, the magnetic signatures at the
transition strongly differ. This is illustrated in Fig. 1b in which the FM
compound shows a dramatic decrease in magnetization as the transition is
crossed (80\% of the value without biasing).

This huge current driven effect can, by no means, be ascribed to a local
magnetic and/or electrical transition as proposed by the filamentary
picture. It seems that, in the case of the FM ground state, the nature of
the interaction between current and moment is much more complex.\ The
experimental observations strongly resemble spintronic features, a kind of
spin-valve effect acting in the reverse.\ An analogy with spin transfer
effect, governed by local exchange, as in magnetic multilayer devices, can
provide an interesting clue to account for experimental data in our
ferromagnetic crystal.\cite{BER96,SLO96,SLO99,BAZ98,HEI01}

A large number of experiments on magnetic elements and multilayers indicate
that the spin of the conduction electrons influences the magnetization of
the elements.\cite{TSO98,SUN99,KAT00,MYE00} In addition, theoretical studies
indicate that spin-polarized current affects the magnetic state of
ferromagnetic conductors via the transfer of angular momentum between the
carriers spins and the conductor magnetic moment.\cite{SLO96,SLO99,BAZ98}
Such interaction would create a so-called ''spin transfer'' torque.\ Most of
the experimental and theoretical works have treated this issue by
considering microdevices having well-controlled geometry. This facilitates a
quantitative study of the spin-transfer effect and allows the testing of the
theoretical models that describe this phenomenon.\ It is not clear how a
nanomagnetic model can account for experimental data in a bulk ferromagnet,
since the magnetization that we measure in our experiment is the average
moment of the entire sample along the direction of the applied field (See
Fig. 2). To be more precise, the ''spintronic effects'' exist when the mean
free path of the polarized electrons is larger than the size of the magnetic
domains.\ The mean free path being in the range of 100 nm, this gives an
upper limit for the typical size of the domains in Fig. 2. In absence of
current in such samples, neutron diffraction experiments\cite{NDE} have
shown that the domain size is larger than 200 nm in $\Pr_{0.8}$Ca$_{0.2}$MnO$%
_3$. This suggests that the current itself would create the nanostructure
necessary for the occurrence of the current-induced magnetization decrease.

Bazaily $et$ $al.$\cite{BAZ98} have generalized the Landau-Lifshitz equation
for a continuously changing magnetization in the presence of a spin
polarized current. This situation is more likely what happens in bulk
ferromagnets where the local moments of adjacent domains (separated by Bloch
walls) have different orientations (provided the applied magnetic field is
well below the saturation field). One reason that the spin-polarized
transport effect should be significant in the perovskite manganites is the
high degree of spin-polarization in these materials; this constitutes the
basis for the double exchange mechanism governing their magnetic ordering.\
In the manganite oxide perovskites, the conduction bandwidth of the $3d$
electrons is likely to be smaller than the Hund coupling energy\cite
{COE95,HWA96}; hence, the carriers in the ferromagnetic ground state are
almost spin polarized, in contrast to the case of the itinerant ferromagnets.

By considering the time-dependent solutions to their general approach,
Bazaily et $al.$\cite{BAZ98} predict a spin wave instability. Other papers
have predicted and observed such a current-induced spin wave generation.\cite
{SLO99,HEI01,MYE00,REZ00} To be excited, spin waves must overcome
anisotropy, exchange and damping effects ; it is found that the current
alters the energy gap of the spin waves for large enough current. In a
recent paper dealing with the same ferromagnetic compound, the amplitude of
the spin waves was studied.\cite{WAHL02} In the latter paper, the amplitude
of the spin waves is controlled by the temperature of the sample and not by
a stream of charge particles. However, the soft nature of the spin waves in $%
\Pr_{0.8}$Ca$_{0.2}$MnO$_3$ is found to be a general trend in the low doped
region of the PCMO system ($D\approx 15\pm 3\ meV.\stackrel{\circ }{A}^2$).
Since generation of spin waves in a conducting ferromagnet requires a
decrease of magnetization, the experimental observation in the FM $\Pr_{0.8}$%
Ca$_{0.2}$MnO$_3$ might be understood by considering the excitation of
spin-waves through the spin transfer framework.\ Moreover, it appears that
this would cause the same observed voltage-current anomalies. As observed in
Fig. 3, the magnetization versus current curves are nonhysteretic and
symmetric. According to Myers et $al.$\cite{MYE00}, nonhysteretic features
are also consistent with spin waves excitations induced by spin transfer.\
It should be noted that the spin transfer theory predicts that the current
necessary to generate spin-wave excitations should increase with field. We
only observe a slight shift of the current threshold with field, which is
not consistent with the expected field dependence. This suggests that the
current necessary to trigger spin wave instability would be rather
independent of the applied magnetic field.\ One might imagine that the
applied magnetic field can eventually re-orient the domains within each
other but fails to make stiffer the spin system inside a domain. Although
all the details of the understanding are not sorted out, the manganese
oxides seem to be of great interest since the magnetic and electric states
are easily tunable by changing the bias current.

Figure Captions

Figure 1 : a V-I characteristic and magnetization versus current density at
100K under 100G for the CO-AF compound.

b : V-I characteristic and magnetization versus current density at 100K
under 100G for the FM compound.

Figure 2 : A current spin polarized J will exert a torque on the domain
moments (m$_1$, m$_2$, m$_3$, ... m$_i$) . For J 
\mbox{$>$}%
J$_{th}$, the interaction of the spin-polarized electrons with the local
moments leads to a deflection of magnetization. As emphasized in the text,
the measured magnetization (M$_{data})$ is the projection of the average
moment of the sample is the direction of the applied field. The hatched area
represents the interface between domains.

Figure 3 : Magnetization versus current density at 100K under 100G for
increasing / decreasing current and for different signs of current bias.

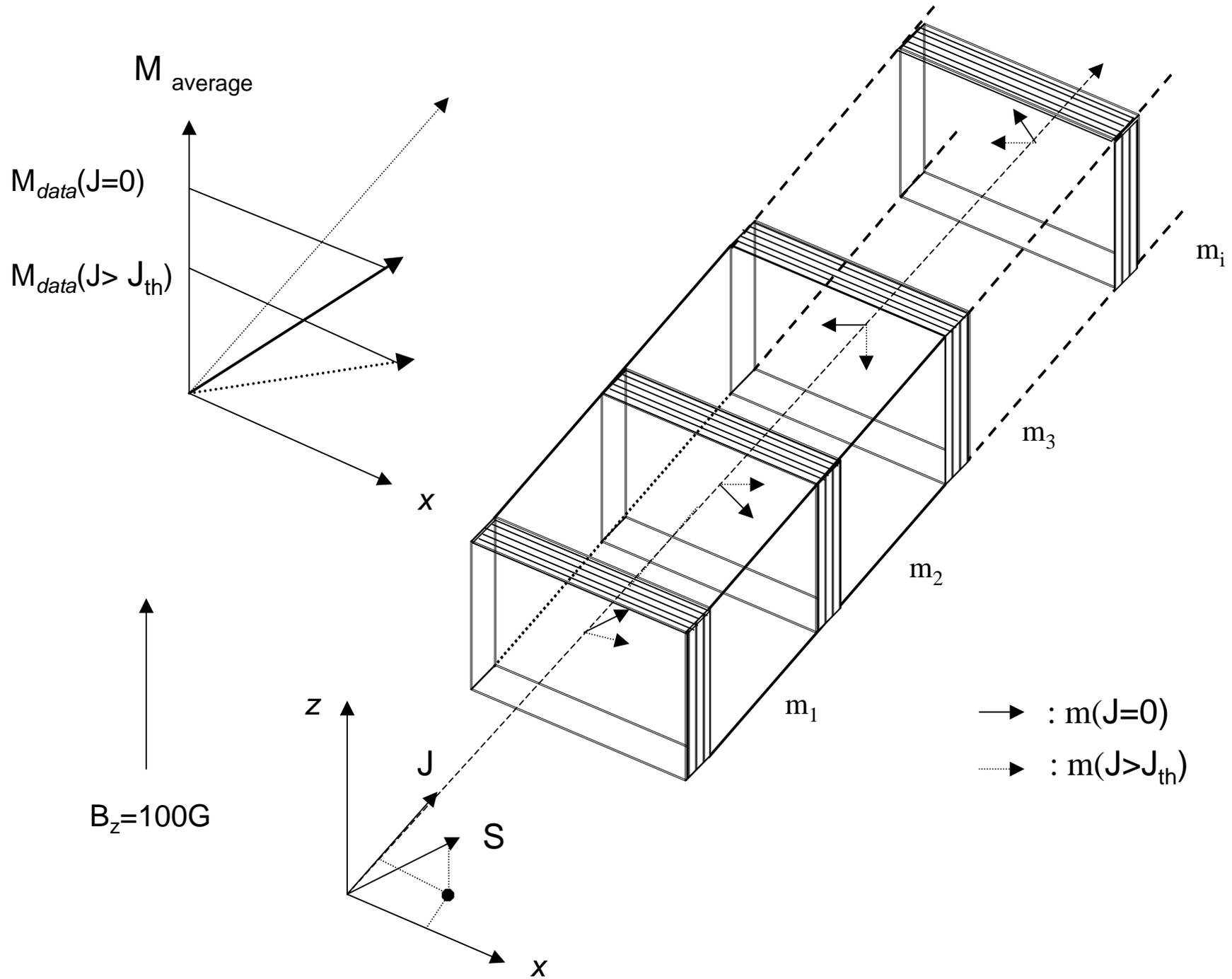
\end{document}